\documentclass[letterpaper]{ptephy} 

\usepackage{amsmath} 
\usepackage{amsthm} 
\usepackage{cite} 

\def\vector#1{\mbox{\boldmath $#1$}}

\begin{document}

\title{Ground-state energies and charge radii of $^{4}$He, $^{16}$O,
$^{40}$Ca, and $^{56}$Ni in
the unitary-model-operator approach}

\author{\name{Takayuki Miyagi}{1*}, \name{Takashi Abe}{1}, \name{Ryoji
Okamoto}{2}, and \name{Takaharu Otsuka}{1,3,4}}


\address{\affil{1}{Department of Physics, the University of Tokyo,
Hongo, Tokyo 113-0033, Japan}
\affil{2}{Senior Academy, Kyushu Institute of Technology, Tobata,
Kitakyushu 804-0015, Japan}
\affil{3}{Center for Nuclear Study, the University of Tokyo, Hongo,
Tokyo 113-0033, Japan}
\affil{4}{National Superconducting Cyclotron Laboratory, Michigan State
University, East Lansing, Michigan 48824, USA}
\email{miyagi@nt.phys.s.u-tokyo.ac.jp}}

\begin{abstract}%
We study the nuclear ground-state properties by using the
 unitary-model-operator approach (UMOA).
Recently, the particle-basis formalism has been introduced in the UMOA and enables us to employ the
 charge-dependent nucleon-nucleon interaction.
We evaluate the ground-state energies
 and charge radii of $^{4}$He, $^{16}$O,
 $^{40}$Ca, and $^{56}$Ni with the charge-dependent Bonn potential.
The ground-state energy is dominated by the contributions from 
 the one- and two-body cluster
 terms, while, for the radius, the one-particle-one-hole excitations
 are more important than the two-particle-two-hole excitations.
The calculated results reproduce the trend of experimental data of the
 saturation property for finite nuclei.
\end{abstract}

\subjectindex{D10}

\maketitle

\section{Introduction\label{sec:intro}}
In nuclear theory, one of the most fundamental problems is to
 describe and understand the nuclear structure
 based on nuclear forces. 
This attempt
 has been made possible owing to the progress of the computational power and 
 that of nuclear many-body approaches. 
For light nuclei with the mass number around $4 \leq A \leq 12$, 
 the {\it ab initio} methods such as the no-core shell model (NCSM) \cite{Barrett:2013} and 
 the Green's function Monte Carlo method \cite{Pieper:2005} have been
 applied extensively to the ground- and excited-state properties. 
However, it is difficult to apply these methods to heavier nuclei because of the
 current computational limitations. 
In order to investigate the medium-mass region, one can apply
 the many-body methods such as the the coupled-cluster method (CCM) \cite{Hagen:2014}
 and the unitary-model-operator approach (UMOA) \cite{SuzukiOkamoto:1994}.
By employing these methods, ground-state energies of doubly magic nuclei
 have been calculated \cite{Heisenberg:1999,Hagen:2007,Hagen:2010,Binder:2013,Fujii:2004,Fujii:2009}.
In the CCM, the systematic calculations have been performed for the ground- and
 excited-state energies around oxygen \cite{Hagen:2012-1} and
 calcium \cite{Hagen:2012-2}.
Such studies have been also done in the self-consistent Green's function method
 \cite{Cipollone:2013,Soma:2014} and the in-medium similarity
 renormalization 
 group approach
 \cite{Tsukiyama:2011,Hergert:2013-1,Hergert:2013-2}.

Besides the nuclear many-body approaches, one of the other important
 ingredients of numerical calculations is the nuclear interaction or potential.
Recently, modern nucleon-nucleon (NN) potentials
 \cite{Stoks:1994,Wiringa:1995,Machleidt:2001,Entem:2003,Epelbaum:2005} have
 been developed and can reproduce with high precision the proton-proton and proton-neutron
 scattering phase shifts, as well as the deuteron properties.
It is known that these NN potentials generally have non-perturbative properties
 caused by the strong repulsive core and the
 singularity coming from the tensor force at short distances.
Therefore, it is difficult to apply such bare potentials directly to 
 the nuclear structure calculations because
 they require the huge model
 space spanned by single-particle basis states.
In order to tailor these short-range correlations to be computationally
 tractable, the effective-interaction theory is generally introduced. 
In the present work, we use the UMOA,
 which is one of the energy-independent Hermitian effective-interaction
 theories.
The calculations of ground-state energies and charge radii
under the isospin symmetry 
 were done for $^{16}$O and $^{40}$Ca by the UMOA
 \cite{SuzukiOkamoto:1987,SuzukiOkamoto:1994,Kumagai:1997}.
Recently, the particle-basis (proton-neutron-basis) formalism was introduced in the UMOA
 \cite{Fujii:2004}.
Since the protons and neutrons are treated explicitly as different
 particles in the particle-basis formalism, charge-dependent NN potentials can be used.
The ground-state energies of $^{16}$O, $^{40}$Ca, and $^{56}$Ni are
 discussed by using such a formalism in Ref. \cite{Fujii:2009}.
However, the charge radius, which is indispensable for understanding of
 the nuclear saturation property, has not been calculated with the 
 charge-dependent framework in the UMOA.
Nowadays, charge radii of many nuclei across the nuclear chart have been
 measured through the elastic electron
 scattering or the isotope shift \cite{Angeli:2013}.
To investigate the charge radius, it is, in principle, desirable to take
 into account the charge-dependence of the NN interaction by the
 particle-basis formalism.

Moreover, the studies of the infinite nuclear matter system at high densities,
 especially the equation of state, are necessary to
 investigate the astrophysical objects such as neutron stars and supernova explosions.
To obtain the equation of state, one of the important issues is to satisfy the
 saturation property of the infinite nuclear matter. 
However, it is difficult to reproduce microscopically the saturation property of such
 system, at least without use of genuine three-nucleon forces.
In order to 
 obtain more reliable predictions for the infinite nuclear matter system,
 one can start from the reproduction of the properties of
 finite nuclei, because the experimental data are available only for finite
 nuclei.
The saturation property of $^{16}$O and $^{40}$Ca was investigated with
 Bonn potentials in the Br\"uckner-Hartree-Fock (BHF) method, and 
 the influence of effective masses of mesons were discussed \cite{Schmid:1991}.
Furthermore, the relativistic effect on $^{16}$O and $^{40}$Ca was
 included through the
 Dirac-Br\"uckner-Hartree-Fock method \cite{Fritz:1993} and improves the
 BHF results.
On the other hand, it is shown that the
 results for such nuclei in the UMOA
 are closer to the experimental values than the BHF results
 \cite{SuzukiOkamoto:1994,Kumagai:1997}
 even without relativistic effect.
Before discussing quantitatively the various contributions, for
 instance, from the relativistic effects and three-body forces, in this
 study, we employ the CD-Bonn potential \cite{Machleidt:2001}, which is
 one of the modern high-precision NN interactions and holds the
 charge-symmetry and charge-independence breakings.
In this Letter, we calculate the ground-state energies 
 and charge radii, corresponding to the inverse of density, of $^{4}$He,
 $^{16}$O, $^{40}$Ca, and $^{56}$Ni, 
 and discuss the saturation property of finite nuclei.


\section{Framework\label{sec:frm}}
Here, we briefly present the calculation method of the ground-state
 energy and charge radius in the 
 UMOA with the particle-basis formalism.
The details can be found, for example, in
Refs. \cite{SuzukiOkamoto:1987,SuzukiOkamoto:1986,SuzukiOkamoto:1994}.
The essence of the UMOA is to construct the effective Hamiltonians
 from the original total Hamiltonian $H$ and
 to obtain the physical observables with effective operators.
The transformed total Hamiltonian, $\widetilde{H}=e^{-S}He^{S}$, is constructed by the unitary operator
 $e^{S}$, and takes into account two-particle-two-hole (2p2h)
 excitations considered as the most important correlations in the nuclear structure.
The exponent $S$ is called as the two-body correlation operator.
Since $e^{S}$ is the many-body operator, $\widetilde{H}$ becomes the
many-body operator.
The transformed Hamiltonian can be decomposed into
 $\widetilde{H}=\widetilde{H}^{(1)}+\widetilde{H}^{(2)}+\widetilde{H}^{(3)}+\cdots$,
 according to the number of
 interacting particles.
Here, $\widetilde{H}^{(1)}$, $\widetilde{H}^{(2)}$, and
 $\widetilde{H}^{(3)}$ are the one-, two-, and three-body cluster terms,
 respectively.
In the practical calculation, we treat the one- and two-body cluster terms directly
 and obtain $S$ by solving the decoupling equation,
 $Q\widetilde{H}^{(2)}P=0$.
Here, $P$ and $Q$ are the projection operators onto the 0p0h and 2p2h
 states, respectively.
Note that the effective interaction is determined self-consistently with
 the one-body potential.

To take the other important correlations into account,
 we diagonalize the transformed Hamiltonian in the 0p0h and 1p1h
 space. 
This procedure is needed because the one-body cluster term induces the
 1p1h excitations.
After the diagonalization, we have the energy of the one- and two-body
 cluster terms, $E^{1+2BC}$.
Due to the diagonalization, the ground state is expressed
 as the linear combination of the unitary transformed wave functions in
 the 0p0h and 1p1h space, 
 $|\Psi\> =
 e^{S}(g_{0}+\sum_{\alpha\beta}
 g_{\alpha\beta}a^{\dag}_{\alpha}b^{\dag}_{\beta})|\phi_{0}\>$.
Here, $|\phi_{0}\>$ is the particle-hole vacuum state, and $g_{0}$ and
 $g_{\alpha\beta}$ are
 coefficients of the 0p0h (particle-hole vacuum) and 1p1h states, respectively. 
Subscripts $\alpha$, $\beta, \cdots$ denote the set of harmonic-oscillator quantum numbers.
The $a^{\dag}_{\alpha}$ and $b^{\dag}_{\beta}$ are creation
 operators of the particle state $\alpha$ and the hole state $\beta$, respectively.
In addition to $E^{1+2BC}$, we evaluate the contribution of the
 three-body cluster term to the ground-state energy, $E^{3BC}$, through
 order $S^{2}$
 \cite{SuzukiOkamoto:1987,SuzukiOkamoto:1986,SuzukiOkamoto:1994} and
 obtain the total energy, $E_{\text{g.s.}}=E^{1+2BC}+E^{3BC}$.

The charge radius $R_{c}$ is evaluated by using the equation,
 $R_{c}^{2} = \langle \Psi|r^{2}|\Psi\rangle + R_{p}^{2}+ (N/Z) R_{n}^{2}$ \cite{Friar:1975}. 
Here, $R_{p}^{2}$ and $R_{n}^{2}$ are the squared radius obtained from the
 experimental charge distribution of the proton and
 neutron, respectively.
Here, we take 
$R_{p}^{2}=0.832$ fm$^{2}$
 \cite{Borisyuk:2010} and 
 $R_{n}^{2}=-0.115$ fm$^{2}$ \cite{Angeli:2013}.
The $Z$ and $N$ are the proton and neutron number, respectively. 
What we calculate in the UMOA is 
 $\langle \Psi|r^{2}|\Psi\rangle$ 
 defined by $\sum_{i}\langle\Psi|(\vector{r}_{i}-\vector{R}_{\text{c.m.}}
)^{2}|\Psi\rangle /Z$.
The $\vector{r}_{i}$ and 
$\vector{R}_{\text{c.m.}}$
are the
 coordinate vectors of the $i$th proton and the center-of-mass (c.m.)
 of 
$A$ nucleons,
respectively.
For $N=Z$ nuclei, 
$\vector{R}_{\text{c.m.}}$ is
 approximately equal to the c.m. coordinate vector of protons,
whose expectation value is approximated as $0s$
 state, i.e., $\langle\Psi|\vector{R}^{2}_{\text{c.m.}}|\Psi\rangle 
 \simeq 
 3\hbar/(2Am\omega)$.
Then, the term $\sum_{i}\langle\Psi|\vector{r}^{2}_{i}|\Psi\rangle/Z$ is 
 expanded up to order $S^{2}$ into
\begin{align}
 \langle\Psi|\sum_{i=1}^{Z} \vector{r}^{2}_{i}|\Psi\rangle& \simeq 
g_{0}^{2}\sum_{\alpha'\le\rho_{F}}r_{\alpha'\alpha'}
 +\frac{g_{0}^{2}}{2} \sum_{\substack{\alpha\beta\gamma >
 \rho_{F}\\ \alpha'\beta' \le \rho_{F}}}
 S_{\alpha\beta\alpha'\beta'}S_{\gamma\beta\alpha'\beta'}r_{\alpha\gamma}
 - \frac{g_{0}^{2}}{2}\sum_{\substack{\alpha\beta >
 \rho_{F}\\ \alpha'\beta'\gamma'
 \le \rho_{F}}}
 S_{\alpha\beta
 \alpha'\beta'}S_{\alpha\beta\gamma'\beta'}r_{\alpha'\gamma'}\notag \\
\label{rccal}
&\hspace{1em}-2g_{0}\sum_{\substack{\alpha >\rho_{F} \\ \alpha'\le
 \rho_{F}}} g_{\alpha\alpha'}
r_{\alpha\alpha'} + 2g_{0}\sum_{\substack{\alpha\beta > \rho_{F} \\
\alpha'\beta'\le
 \rho_{F}}}g_{\beta\beta'}S_{\alpha\beta\alpha'\beta'}r_{\alpha\alpha'}f_{z_{\beta'}}
 \\
&\hspace{1em}+\sum_{\substack{\alpha > \rho_{F} \\ \alpha'\beta'\le \rho_{F}}}
g_{\alpha\alpha'}g_{\alpha\alpha'}r_{\beta'\beta'}
 +\sum_{\substack{\alpha\beta > \rho_{F} \\ \alpha'\le
 \rho_{F}}}g_{\alpha\alpha'} g_{\beta\alpha'}r_{\alpha\beta}
 - \sum_{\substack{\alpha >\rho_{F} \\ \alpha'\beta'\le \rho_{F}}}
 g_{\alpha\alpha'}g_{\alpha\beta'}
 r_{\alpha'\beta'}. \notag
\end{align}
Here, we use the notations,
 $r_{\alpha\beta}=\langle\alpha|\vector{r}^{2}|\beta\rangle$ and 
 $S_{\alpha\beta\gamma\delta}=\langle\alpha\beta|S|\gamma\delta\rangle$.
The $\rho_{F}$ denotes the Fermi level.
The matrix element of $r^{2}$ is evaluated only with
 respect to the proton states.
Note that summations in Eq. (\ref{rccal}) run 
 over only the proton states. 
However, indices $\beta$
 and $\beta'$ in the second, third, and fifth terms in
 Eq. (\ref{rccal}) run over both of the proton and neutron states.
The isospin factor $f_{z_{\alpha'}}$ is $-1$ $(+1)$ for proton (neutron).

\section{Results and discussions\label{sec:res}}

\begin{table}[t]
\begin{center}
\caption{Ground-state energy for each nucleus. 
 We take $\rho_{1}=12$ and $\hbar\omega_{\text{min}}=18$ MeV for $^{4}$He, 
$\rho_{1}=14$ and $\hbar\omega_{\text{min}}=15$
 MeV for $^{16}$O, $\rho_{1}=18$ and $\hbar\omega_\text{min}=14$ MeV for
 $^{40}$Ca, and $\rho_{1}=20$ and $\hbar\omega_{\text{min}}=14$ MeV for
 $^{56}$Ni. 
The definitions of $E^{1+2BC}$, $E^{3BC}$, and $E_{\text{g.s.}}$ are
 given in the text.
The experimental values are taken from
 Ref.\cite{Wang:2012}. 
All the energies are in units of MeV.}
{\tabcolsep=8mm
  \begin{tabular}{lrrrr} \hline\hline
          &$^{4}$He & $^{16}$O & $^{40}$Ca & $^{56}$Ni\\ \hline 
    $E^{1+2BC}$&-26.13 & -115.58 & -334.36& -454.84 \\
    $E^{3BC}$& -1.60&-3.82 & -5.92& -18.20 \\
    $E_{\text{g.s.}}$&-27.73 & -119.39 & -340.28& -473.04\\
    Expt. &-28.30 & -127.62 & -342.05& -483.99\\ \hline \hline
  \end{tabular}}
\label{T:Egs}
\end{center}
\end{table}

In this section, we show the results of the ground-state energy
 and charge radius of $^{4}$He, $^{16}$O, $^{40}$Ca, and $^{56}$Ni, and
 discuss the saturation property of these nuclei. 
All the calculated results are obtained with the CD-Bonn potential \cite{Machleidt:2001}. 
The Coulomb interaction is included in the proton-proton channel.
Following Refs. \cite{Fujii:2004,Fujii:2009}, we adopt the two-step
 decoupling method.
At the second-step decoupling, we add the
 term, $H_{\text{c.m.}} = \beta(T_{\text{c.m.}} + U_{\text{c.m.}} -
 3\hbar\omega/2)$, to the intrinsic Hamiltonian $\widetilde{H}-T_{\text{c.m.}}$
 so as to remove the c.m. spurious
 motion by the Gl\"ockner-Lawson prescription with $\beta=1$.
Here, $T_{\text{c.m.}}$ and $U_{\text{c.m.}}$ are the kinetic and harmonic oscillator
 potential of c.m., respectively.
The details can be found in Ref. \cite{Fujii:2004}.
The model-space size is defined by the sum of the relevant quantum
 numbers of the two-body state,
 $\rho_{1}=2n_{\alpha}+l_{\alpha}+2n_{\beta}+l_{\beta}$.
Here, $n_{\alpha}$ and $l_{\alpha}$ are the principal and
 azimuthal quantum numbers of the harmonic-oscillator state $\alpha$,
 respectively.
We perform the calculations with various $\rho_{1}$ and the
 harmonic-oscillator energy $\hbar\omega$, and investigate the
 $\rho_{1}$- and $\hbar\omega$-dependence of the
 ground-state energies and charge radii.

We observed that the ground-state energy lowers monotonically as increasing 
 $\rho_{1}$ similar to
 Ref. \cite{Fujii:2009}.
This is non-trivial because our calculations do not have to preserve
 the variational principle mainly due to the truncation of four- and higher-body cluster
 terms.
When many-body cluster terms are truncated, the dependence of the model-space size
 on the energy does not obey the variational principle and can be found, for example, 
 in the case of the NCSM \cite{Navratil:2000} with the
 Lee-Suzuki transformed effective interaction \cite{Lee:1980,Suzuki:1980}.
Actually, our ground-state energy of $^{4}$He is overbound 
 a little compared with the NCSM result in the sufficiently large model space, 
 which is thought of as the exact solution with the CD-Bonn potential.
Our final results of the ground-state energy depend slightly on
 $\hbar\omega$ even if the $\rho_{1}$-dependence vanishes.
Since the choice of $\hbar\omega$ is arbitrary in nature, the $\hbar\omega$-dependence 
 should vanish at sufficiently large $\rho_{1}$.
In the case of the CCM, the $\hbar\omega$-dependence at
 the coupled-cluster double (CCD) level, similar to our framework, vanishes
 at the coupled-cluster single and double (CCSD) level \cite{Kohno:2012}.
Therefore, the $\hbar\omega$-dependence in the UMOA could be reduced by
 introducing the one-body correlation operator, in addition to the
 two-body correlation operator. 
It was also shown that 
 the ground-state energy at 
$\hbar\omega_{\text{min}}$
 in the CCD is
 close to the result in the CCSD.
Here, $\hbar\omega_{\text{min}}$ is the value of $\hbar\omega$
 minimizing the ground-state energy.
Thus, we tabulate the ground-state energy of each nucleus
 at 
$\hbar\omega_{\text{min}}$
in Table \ref{T:Egs}.
The $E^{1+2BC}$ is the energy obtained from the one- and two-body cluster terms.
The contribution of the three-body cluster term, 
 $E^{3BC}$, is much smaller than $E^{1+2BC}$ and
 attractive for all nuclei examined here. 
This tendency is also observed in the calculated results obtained by the 
 CCM \cite{Hagen:2007,Hagen:2010}.
The contributions from the higher-body cluster terms would be less than those
 of the three-body cluster term.
Thus the ground-state energies, $E_{\text{g.s.}}=E^{1+2BC}+E^{3BC}$, are
 expected to almost converge with respect to the cluster expansion.

\begin{table}[t]
\begin{center}
\caption{Charge radius $R_{c}$
 for each nucleus. 
The entries of ``$R_{c}$ w/o 1p1h and 2p2h correlations'', ``$R_{c}$ w/o 1p1h correlations'', and
 ``$R_{c}$ w/o 2p2h correlations'' are the results from  Eq. (\ref{rccal})
 with both of $S_{\alpha\beta\alpha'\beta'}=0$
 and $g_{\alpha\alpha'}=0$, $g_{\alpha\alpha'}=0$, and 
 $S_{\alpha\beta\alpha'\beta'}=0$, respectively.
The model-space size and $\hbar\omega_{\text{min}}$ are same as in
 Table \ref{T:Egs}.
The experimental values are taken from
 Ref.\cite{Angeli:2013}. All the radii are in units of fm.
}
{\tabcolsep=5mm
   \begin{tabular}{lrrrr} \hline\hline
         & $^{4}$He & $^{16}$O & $^{40}$Ca & $^{56}$Ni\\ \hline
   $R_{c}$ w/o 1p1h and 2p2h correlations &1.81   & 2.59    &3.08 & 3.28  \\ 
   $R_{c}$ w/o 1p1h correlations&1.82  &2.60   & 3.09 & 3.29 \\
   $R_{c}$ w/o 2p2h correlations&1.67  &2.44   & 2.97 & 3.20 \\
   $R_{c}$ &1.67  &2.44   &  2.97 &3.19  \\
    Expt. &1.68 &2.69 & 3.48&  \\ \hline \hline
  \end{tabular}}
\label{T:Rc}
\end{center}
\end{table}

In contrast to the ground-state energies, charge radii
 have the sizable $\hbar\omega$-dependence
 except for $^{4}$He even if the results converge with respect to
 $\rho_{1}$.
The charge radii decrease monotonically as $\hbar\omega$
 increases.
 For example, the charge radius of $^{16}$O ($^{56}$Ni) changes from 2.76
 (3.56) fm at $\hbar\omega = 11$ MeV to 2.27 (2.87) fm at $\hbar\omega =
 20$ MeV.
To evaluate the charge radius in the UMOA, $\hbar\omega$ was formerly
 taken to 
$\hbar\omega_{\text{min}}$ \cite{SuzukiOkamoto:1994,Kumagai:1997}.
Recently, it was demonstrated that the radius at 
$\hbar\omega_{\text{min}}$
 coincides the
 $\hbar\omega$-independent result in the CCM \cite{Kohno:2012}.
Therefore, in Table \ref{T:Rc}, the charge radius of each nucleus at 
$\hbar\omega_{\text{min}}$
 is tabulated. 
The contributions of the correlation operator $S$ and the diagonalization
 coefficients $g_{\alpha\alpha'}$ are also shown as those of 2p2h and
 1p1h correlations, respectively.
Since the effects of the correlation operator $S$ are much smaller than those of
 $g_{\alpha\alpha'}$, 1p1h excitations to the charge
 radius are more important than 2p2h excitations.
In other words, for the radius, it can be expected that the consideration of the one-body
 correlation operator is more important than that of the
 two-body correlation operator.
All the charge radii investigated here shrink compared with the
 experimental values as increasing the mass number.
Note that the charge radius of $^{56}$Ni has not been measured yet.
In our framework, neutron and matter radii can be also calculated in
 the same way.
The differences between the proton and neutron radii for the nuclei
examined here are smaller than 0.02 fm. Therefore, we only discuss the
proton radius in this work.

\begin{figure}[t]
\begin{center}
\centerline{
\includegraphics[width=4in,height=2.6in,clip]{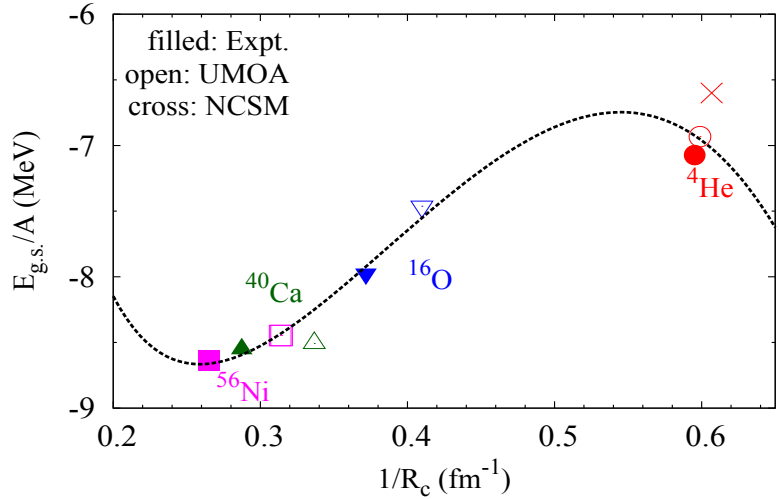}
\label{F:satCa}}
\caption{Ground-state energy per nucleon and inversed charge radius of
 $^{4}$He, $^{16}$O, $^{40}$Ca, and $^{56}$Ni. 
Filled symbols correspond
 to the experimental values.
As for the charge radius of $^{56}$Ni, we substitute the experimental value of $^{58}$Ni.
Open symbols correspond to the numerical results by the UMOA.
The cross symbol is the NCSM result \cite{Navratil:2000}.
The circles, down triangles, up triangles, and squares are for $^{4}$He,
 $^{16}$O, $^{40}$Ca, and $^{56}$Ni, respectively.
The dashed curve is obtained from the Bethe-Weizs\"acker
 formula for the energy and
 $R_{c}=r_{0}A^{-1/3}$ for the radius \cite{Ring:1980,Henley:2007}. 
 The adopted parameters are shown in the text.}
\label{F:sat}
\end{center}
\vspace{-2em}
\end{figure}

In Fig. \ref{F:sat}, the 
 saturation property of $^{4}$He, $^{16}$O,
 $^{40}$Ca, and $^{56}$Ni is illustrated.
Our calculated results of $^{16}$O and $^{40}$Ca are consistent with
 the results by the earlier UMOA calculations, which were obtained with
 the realistic NN potentials at 
$\hbar\omega_{\text{min}}$
 \cite{SuzukiOkamoto:1994,Kumagai:1997}.
The dashed curve, which is obtained by the empirical formula given by Bethe and
 Weizs\"acker \cite{Ring:1980},
 $E/A=a_{\text{V}}+a_{\text{S}}A^{-1/3}+a_{\text{C}}Z^{2}/A^{4/3}+a_{\text{I}}(N-Z)^{2}/A^{2}
 -a_{\text{P}}A^{-7/4}$, 
 and the charge radius, $R_{c}=r_{0}A^{1/3}$ 
\cite{Ring:1980,Henley:2007}
 shows the systematic behavior for light- and medium-mass $N=Z$ nuclei.
Here, we adjust $a_{\text{V}}=-16.10$ MeV and $r_{0}=1.05$ fm, and
 use $a_{\text{S}}=18.56$ MeV,
 $a_{\text{C}}=0.717$ MeV, $a_{\text{I}}=28.1$ MeV, and $a_{\text{P}} =
 34.0$ MeV as found in Ref. \cite{Ring:1980}.
Our calculated results show similar pattern to this empirical curve.
However, taking a closer look at our results with the NN interaction
 only, all the binding energies are underbound to and also all the charge
 radii are smaller than the experimental data.
When we use the other realistic NN potentials, 
 the binding energies are expected to increase (decrease) with decreasing
 (increasing) charge radii \cite{Schmid:1991}, and
 distribute over the Coester line \cite{Ring:1980}.
Actually, the numerical results by the UMOA given in Refs. \cite{SuzukiOkamoto:1994,Kumagai:1997},
 show such a correlation between the binding energy and charge radius.
Therefore, it is likely that the results do not approach the experimental
 data even if we use other modern high-precision NN interactions.
The deviation from the experimental data might come from the 
 lack of the genuine three-nucleon force effect, because it gives
 the attraction in light nuclei as shown in the
 Green's function Monte Carlo method
 \cite{Pieper:2001}.
Also, the three-nucleon force through the relativistic framework stretches the nuclei as found from the
 comparison between the Br\"uckner-Hartree-Fock and
 Dirac-Br\"uckner-Hartree-Fock results \cite{Fritz:1993}.
Combining these facts, the three-nucleon force seems to be necessary to
 reproduce the binding energy and charge radius simultaneously.
From the recent {\it ab initio} calculation of Green's function theory
 \cite{Barbieri:2012}, the three-nucleon force effect actually increases
 the matter radii of $^{16}$O and $^{44}$Ca.
However, the discrepancy between the experimental data and the recent {\it ab
 initio} results still remains and needs to be further investigations
 \cite{Soma:2014,Binder:2014}.
As for $^{4}$He, we can compare with the result obtained by the NCSM
 \cite{Navratil:2000} which is also plotted in Fig
 \ref{F:sat}. 
Our result is close to the NCSM result.
The NCSM result can be considered as the exact solution with the CD-Bonn
 potential, because the calculation was performed at sufficiently large
 model space where the $\hbar\omega$- and model-space size dependencies
 are negligible. 
Our charge radius for $^{4}$He can be determined uniquely, because of the weak
 $\hbar\omega$-dependence.
For $^{16}$O, $^{40}$Ca, and $^{56}$Ni, charge radii have sizable $\hbar\omega$-dependence.
Thus, we cannot discuss these results in the same accuracy as
 $^{4}$He. 
It is necessary to obtain the practically $\hbar\omega$-independent results.
To achieve this, 
 the consideration of the one-body correlation operator would be desirable,
 i.e. the construction of the effective Hamiltonian which does not
 induce the 1p1h excitations in addition to the 2p2h excitations by the unitary transformation.

\section{Summary\label{sec:con}}
We represent the way to obtain the ground-state energy and charge radius in
 the particle-basis formalism of the UMOA. 
We evaluate the ground-state energy and charge radius for
 $^{4}$He, $^{16}$O, $^{40}$Ca, and $^{56}$Ni with the CD-Bonn potential.
The results are obtained in the model space
 defined by the two-body state, $\rho_{1}=2n_{\alpha}+l_{\alpha}+2n_{\beta}+l_{\beta}$,
 and converge with respect to $\rho_{1}$.
We find almost converged ground-state energies with respect to the
 cluster expansion, because the contribution of the three-body cluster
 term estimated here is much smaller than the one- and two-body cluster
 terms. 
All the ground-state energies calculated here are underbound to the experimental data.
The contribution of the two-body correlation operator $S$ to the charge radius is 
 less important than 1p1h excitations.
All the charge radii estimated here are smaller than the experimental data.
From our results of the UMOA, we show the obtained saturation property of
 finite nuclei, consistently with the trend of experimental data.
Our result of $^{4}$He is close to the {\it ab initio}
 solution with the same interaction.

In this Letter, we take
 $\hbar\omega$ minimizing the ground-state energy. 
The validity of this choice is discussed in the CCM \cite{Kohno:2012}.
However, it is necessary to obtain the virtually $\hbar\omega$-independent results for the
 quantitative comparison with the experimental data.
In the CCM, the role of
 the one-body cluster operator is investigated \cite{Kohno:2012}, and
 it is shown to reduce the $\hbar\omega$-dependence.
If we include the one-body correlation operator, 
 in addition to the two-body correlation operator, the
 $\hbar\omega$-dependence is expected to be weakened.
The results in the UMOA with the one-body correlation operator 
 will be reported elsewhere in the near future.
It is also found that the genuine three-nucleon force seems to be
 necessary so as to quantitatively reproduce the saturation property of
 finite nuclei.
The inclusion of the three-body force in the UMOA is under way.

\ack
The authors thank M.~Kohno, K.~Suzuki, H.~Kumagai, S.~Fujii, N.~Shimizu, and B.~R.~Barrett
for many useful discussions.
The part of numerical calculation has been done on a supercomputer (NEC SX8R) at Research
Center for Nuclear Physics, Osaka University.
This work was supported in part by MEXT SPIRE and JICFuS.
It was also supported in part by the Program in part for Leading Graduate Schools,
MEXT, Japan.\\

\begin{thebibliography}{99}
\bibitem{Barrett:2013}
	B.~R.~Barrett, P.~Navr\'atil, and J.~P.~Vary,
	Prog. Part. Nucl. Phys. {\bf 69}, 131 (2013) and references therein.
\bibitem{Pieper:2005}
	S.~C.~Pieper and R.~B.~Wiringa,
	Ann. Rev. Nucl. Part. Sci. {\bf 51}, 53 (2001) and references therein.
\bibitem{Hagen:2014}
	G.~Hagen, T.~Papenbrock, M.~Hjorth-Jensen, and D.~J.~Dean,
	Rept. Prog. Phys. {\bf 77}, 096302 (2014) and references therein.
\bibitem{SuzukiOkamoto:1994}
	K.~Suzuki and R.~Okamoto,
	Prog. Theor. Phys. {\bf 96}, 1045 (1994) and references therein.
\bibitem{Fujii:2004}
	S.~Fujii, R.~Okamoto, and K.~Suzuki, 
	\PRC{69,034328,2004}
\bibitem{Fujii:2009}
	S.~Fujii, R.~Okamoto, and K.~Suzuki,
	\PRL{103,182501,2009}
\bibitem{Heisenberg:1999}
	J.~H.~Heisenberg and B.~Mihaila,
	\PRC{59,1440,1999}
\bibitem{Hagen:2007}
	G.~Hagen, D.~J.~Dean, M.~Hjorth-Jensen, T.~Papenbrock, and
	A.~Schwenk,
	\PRC{76,044305,2007}
\bibitem{Hagen:2010}
	G.~Hagen, T.~Papenbrock, D.~J.~Dean, and M.~Hjorth-Jensen,
	\PRC{82,034330,2010}
\bibitem{Binder:2013}
	S.~Binder, J.~Langhammer, A.~Calci, P.~Navr\'atil, and R.~Roth,
	\PRC{87,021303,2013}
\bibitem{Hagen:2012-1}
	G.~Hagen, M.~Hjorth-Jensen, G.~R.~Jansen, R.~Machleidt, and
	T.~Papenbrock,
	\PRL{108,242501,2012}
\bibitem{Hagen:2012-2}
	G.~Hagen, M.~Hjorth-Jensen, G.~R.~Jansen, R.~Machleidt, and
	T.~Papenbrock,
	\PRL{109,032502,2012}
\bibitem{Cipollone:2013}
	A.~Cipollone, C.~Barbieri, and P.~Navr\'atil,
	\PRL{111,062501,2013}
\bibitem{Soma:2014}
	V.~Som\`a, A.~Cipollone, C.~Barbieri, P.~Navr\'atil, and
	T.~Duguet,
	\PRC{89,061301(R),2014}
\bibitem{Tsukiyama:2011}
	K.~Tsukiyama, S.~K.~Bogner, and A.~Schwenk,
	\PRL{106,222502,2011}
\bibitem{Hergert:2013-1}
	H.~Hergert, S.~K.~Bogner, S.~Binder, A.~Calci, J.~Langhammer,
	R.~Roth, and A.~Schwenk,
	\PRC{87,034307,2013}
\bibitem{Hergert:2013-2}
	H.~Hergert, S.~Binder, A.~Calci, J.~Langhammer, and R.~Roth,
	\PRL{110,242501,2013}

 \bibitem{Stoks:1994}
	 V.~G.~J.~Stoks, R.~A.~M.~Klomp, C.~P.~F.~Terheggen, and J.~J.~de
	 Swart,
	 \PRC{49,2950,1994}
\bibitem{Wiringa:1995}
	R.~B.~Wiringa, V.~G.~J.~Stoks, and R.~Schiavilla
	\PRC{51,38,1995}
 \bibitem{Machleidt:2001}
	 R.~Machleidt, 
	 \PRC{63,024001,2001}
 \bibitem{Entem:2003}
	 D.~R.~Entem and R.~Machliedt,
	 \PRC{68,041001,2003}
 \bibitem{Epelbaum:2005}
	 E.~Epelbaum, W.~Gl\"ockle, and U.-G.~Mei\ss ner,
	 \NPA{747,362,2005}
\bibitem{SuzukiOkamoto:1987}
	K.~Suzuki, R.~Okamoto, and H.~Kumagai, 
	\PRC{36,804,1987}
\bibitem{Kumagai:1997}
	H.~Kumagai, K.~Suzuki, and R.~Okamoto, 
	\PTP{97,1023,1997}
\bibitem{Angeli:2013}
	I.~Angeli and K.~P.~Marinova,
	At. Data Nuc. Data Tables, {\bf 99}, 69 (2013).
\bibitem{Schmid:1991}
	K.~W.~Schmid, H.~M\"uther, and R.~Machleidt,
	\NPA{530,14,1991}
\bibitem{Fritz:1993}
	R.~Fritz, H.~M\"uther, and R.~Machleidt,
	\PRL{71,46,1993}
\bibitem{SuzukiOkamoto:1986}
	K.~Suzuki and R.~Okamoto,
	\PTP{75,1388,1986}
\bibitem{Friar:1975}
	J.~L.~Friar and J.~W.~Negele,
	Adv. Nucl. Phys. {\bf 8}, 219 (1975).
\bibitem{Borisyuk:2010}
	D.~Borisyuk,
	\NPA{843,59,2010}
\bibitem{Navratil:2000}
	P.~Navr\'atil, G.~P.~Kamuntavi\v{c}ius, and B.~R.~Barrett, 
	\PRC{61,044001,2000}
\bibitem{Lee:1980}
	S.~Y.~Lee and K.~Suzuki,
	\PLB{91,173,1980}
\bibitem{Suzuki:1980}
	K.~Suzuki and S.~Y.Lee, 
	\PTP{64,2091,1980}
\bibitem{Wang:2012}
	M.~Wang et al., Chin. Phys., C {\bf 36}, 1603 (2012).
\bibitem{Kohno:2012}
	M.~Kohno and R.~Okamoto,
	\PRC{86,014317,2012}
\bibitem{Ring:1980}
	P.~Ring and P.~Schuck, {\it The Nuclear Many-Body Problem}
	(Springer, Berlin, 1980), 3rd ed.
\bibitem{Henley:2007}
	E.~M.~Henley and A.~Garcia, {\it Subatomic Physics}
	(World Scientific, Singapore, 2007.), 3rd ed.,p144.
\bibitem{Pieper:2001}
	S.~C.~Pieper, V.~R.~Pandharipande, R.~B.~Wiringa, and
	J.~Carlson,
	\PRC{64,014001,2001}
\bibitem{Barbieri:2012}
	C.~Barbieri, A.~Cipollone, V.~Som\`a, T.~Duguet, and
	P.~Navr\'atil,
	arXiv:1211.3315 [nucl-th] (2012). 
\bibitem{Binder:2014}
	S.~Binder, J.~Langhammer, A.~Calci, and R.~Roth,
	\PLB{736,119,2014}
\end{thebibliography}

\appendix
\end{document}